\documentclass[12pt,preprint]{aastex}

\shorttitle{Gravitational lens CFRS03.1077}
\shortauthors{Crampton et al.}

\begin{document}


\title{The Gravitational Lens CFRS03.1077$\footnotemark[1]$ }


\author{David Crampton\altaffilmark{2} and David Schade\altaffilmark{2}} 
\affil{Dominion Astrophysical Observatory, HIA,
National Research Council of Canada,\\ Victoria, B.C. V9E 2E7, Canada}
\email{David.Crampton@nrc.ca}
\author{F. Hammer\altaffilmark{2} and A. Matzkin}   
\affil{Observatoire de Paris, Section de Meudon, DAEC, 92195 Meudon Principal
 Cedex, France. }
\author{S.J. Lilly\altaffilmark{2}}
\affil{Dominion Astrophysical Observatory, HIA,
National Research Council of Canada,\\ Victoria, B.C. V9E 2E7, Canada}
\author{O. Le F\`evre\altaffilmark{2}}
\affil{Laboratoire d'Astronomie Spatiale, 13376 Marseille, France}

\footnotetext[1]{Based on observations with the NASA/ESA {\it Hubble
Space Telescope} obtained at the Space Telescope Science Institute,
which is operated by the Association of Universities for Research in
Astronomy Inc., under NASA contract NAS 5-26555. }

\altaffiltext{2} {Visiting Astronomer, Canada-France-Hawaii Telescope,
operated by the National Research Council of Canada, the Centre de la
Recherche Scientifique de France, and the University of Hawaii}

\begin{abstract}

An exquisite gravitational arc with a radius of 2\farcs1 has been
discovered around the $z = 0.938$ field elliptical galaxy CFRS03.1077
during $HST$ observations of Canada-France Redshift Survey (CFRS)
fields.  Spectroscopic observations of the arc show that the redshift
of the resolved lensed galaxy is z = 2.941.  This gravitational
lens-source system is well-fitted using the position angle and
ellipticity derived from the visible matter distribution and an
isothermal mass profile with a mass corresponding to $\sigma=387\pm5$
km s$^{-1}$. Surprisingly, given the evidence for passive evolution of
elliptical galaxies, this is in good agreement with an estimate based on
the fundamental plane for $z=0$ ellipticals. This, perhaps, indicates
that this galaxy has not shared in the significant evolution observed
for average elliptical galaxies at z$\sim$ 1. A second elliptical galaxy
with similar luminosity from the CFRS survey, CFRS 14.1311 at
$z=0.807$, is also a lens but in this case the lens model
gives a much smaller mass-to-light ratio, i.e., it appears to confirm
the expected evolution.  This suggests that this pair of field
elliptical galaxies may have very different evolutionary histories, a
significant result if confirmed.  Clearly, CFRS03.1077 demonstrates
that these ``Einstein rings" are powerful probes of high redshift
galaxies.

\end{abstract}


\section{INTRODUCTION}

A very beautiful arc structure, characteristic of a lensed object, was
noted around the galaxy CFRS03.1077 \citep{Ham95} during analysis of
$HST$ images of high redshift field galaxies. CFRS03.1077 is a bright
elliptical galaxy at z = 0.938 and hence is likely to lens background
galaxies or quasars. \citet{Mir92} predicted that $\sim$100
gravitationally lensed rings per square degree should be observable at
optical wavelengths to $B \sim 26$, given the number of Einstein rings
at radio wavelengths and the projected density of galaxies in optical
surveys.  Indeed, \citet{Rat99} report finding 10 good galaxy lens
candidates in 0.1 square degrees, based on analyses of $HST$ Medium Deep
Survey images. These, however, have not yet been confirmed and remain
only as candidates.

It is now recognized that galaxies which act as gravitational lenses
are extremely important tools for studying a variety of cosmological
problems including the nature and evolution of the lensing galaxies
themselves \citep{Koc00,Nar99,Ref94, Schn96}. For example, \citet{Im97} showed that observations of seven galaxy lenses  favor
a nonzero cosmological constant, while \citet{Koc00} and
\citet{Kee98} present results on the properties of
the lensing galaxies, most of which are early-type. Since the lens
models give data on the total (dark + visible) distribution of mass,
investigations like the latter are extremely powerful, especially for
studying the properties of galaxies at high redshifts where traditional
methods become increasingly difficult. Einstein ring lenses provide 
additional constraints and hence are even more important to the
determination of mass distributions of the lensing galaxies \citep{Koc01}.

Even though Einstein ring lenses are predicted to be much more common
at optical wavelengths than radio, only one optically-identified system
has so far been confirmed:  0047-2808 \citep{War98, War99}. Rather than
being initially recognised from an image as a ``ring", this lens was
detected through superposition of emission lines from the lensed object
on the spectrum of the foreground galaxy. \citet{Hew00} and
\citet{Hal00} recently report finding additional candidate lenses in
this manner, but these remain to be confirmed. These authors argue that
spectroscopic observation of distant early type galaxies is one of the
best methods of detecting galaxy lenses, particularly because the
presence of spectroscopic features from the source implies that the
crucial redshifts of both the source and deflector can be determined.
In practice, however, this method is only viable for background
galaxies with strong emission lines.  Furthermore, galaxy lens
candidates are quite easily recognizable with images from $HST$ or
ground-based adaptive optic systems, and disentangling the spectral
features arising from the source and deflector galaxies is no longer as
difficult with the improved spatial resolution being achieved on modern
large ground based telescopes.

In this paper we report on a detailed analysis of the $HST$ images of
the lens and source galaxies related to CFRS03.1077. We also present
spectra of these galaxies obtained with the Canada-France-Hawaii
Telescope (CFHT) and derive the redshift of the source - a galaxy
without strong emission lines. Finally, a simple model of the
lens-source system is described.

\section{OBSERVATIONS}

\subsection{$HST$ imaging}

$HST$ WFPC-2 observations were obtained on 1996 January 10 of the
CFRS0300+00 field (Lilly et al. 1995) during a  survey of the
morphology of high redshift field galaxies in CFRS fields.  The total
exposure time for the F814W images of the CFRS0300+00 field was 6700s.
Even though CFRS03.1077 was not located in the higher resolution PC
part of the WFPC-2 field, the arc structure near the galaxy was
immediately obvious while preprocessing the data (see Figure 1a) and
was clear in the individual frames before stacking.

\subsubsection{The Lensing Galaxy}

The J2000 coordinates of CFRS03.1077 are 03$^h$ 02$^m$ 30\fs9
+00\arcdeg\ 06\arcmin\ 02\farcs1.  The arc has a radius of $2\farcs1\pm
0\farcs1$ and is accurately centered (to within $\pm $0\farcs05) on the
galaxy.

Special attention was paid to deriving the parameters of CFRS03.1077.
A high S/N point spread function was determined from six stars located
at similar locations on images of other fields (since no suitable
nearby star was present) and several other fainter stars were used to
assess possible errors.  Fitting was carried out on a ``symmetrized''
image (see \citet{Sch95, Sch96} for details) of the galaxy so that the
procedure was not perturbed by the presence of the arc or any other
images. The best-fit value of the reduced chi-square was 0.9988 for a
fit radius of 20 pixels.  Subtraction of the model galaxy (see Figure
1b) demonstrates that the resulting fit is an extremely good
representation of the galaxy's luminosity distribution.

The parameters of the galaxy were measured from the combined F814W
image with a total integration time of 6700s.  The errors in the
parameters were estimated from the fits to the five individual $HST$
images with exposure times of 1100, 1200, or 2100 s. The dispersions
among these fits were 0\farcs13 in $R_e$, 0.025 in axial ratio (b/a),
2\fdg8 in position angle, 0.04 mag in total magnitude, and 0.06 mag in
central surface brightness. The errors in the stacked image are
expected to be roughly $\sqrt 5$ times smaller because of better
signal-to-noise ratio. These errors are also likely to be
systematically over estimated because cosmic rays are not properly
removed in the individual images. Possible systematic errors due to
imperfect knowledge of the point-spread-function (PSF) were estimated
by repeating the fits with several different PSFs from different
observations and different positions on the chip. The results show that
these systematic errors are a few percent or less in all parameters.
The derived parameters depend weakly on the choice of fitting radius
(the radius over which $\chi^2$ was computed) and these were estimated
from simulations of this galaxy with realistic backgrounds.  For
reasonable values of fitting radius, the errors were smaller than 3\%,
but reached extreme values of 6\% for fitting radii much smaller than
the half-light radius. We adopt errors corresponding to those from the
individual image fits divided by $\sqrt 5$ and then add 5\% in
quadrature to allow for possible systematic errors.  Details of the
parameters of the lensing galaxy are summarized in Table 1.

At this redshift, the lensing galaxy has a half-light radius of
$R_e=14.7\pm 1$ kpc and $M_{AB}(B)=-22.77\pm0.1$ for H$_o=50$ km
s$^{-1}$ Mpc$^{-1}$, q$_o=0.5$. The $V-I$ color of the galaxy is
typical of galaxies with z $\sim$1 in the CFRS sample.  The computed
rest-frame color is $(U-V)_{AB}=1.53$.

\subsubsection{The Arc and counter-images}

The measured properties of the arc image are given in Table 2.  The arc
has a length of 2\farcs1 and is centered on the galaxy to within $\pm
0\farcs05$ (the precision of the measurement). The observed surface
brightness is 21.2 magnitudes per square arcsecond in the brightest
pixel which corresponds to 16.73 in rest-frame AB(2030) magnitudes.
This surface brightness is modulated by pixellation and by the
convolution with the HST point-spread function. For a compact
exponential disk the implied central surface brightness would be about
2 magnitudes brighter than the observed surface brightness and for a
compact $R^{1/4}$ bulge the implied central surface brightness would be
higher by $\sim 7$ magnitudes. The exact values depend on the details
of the luminosity profiles.

The arc is resolved across its narrowest dimension, i.e., it has
non-zero intrinsic width compared with the point-spread function of a
star on the same frame. One dimensional gaussian fits across the arc at
various points along its length give a median observed width (FWHM) of
0\farcs23 and a dispersion of 0\farcs025. For comparison, a fit to a
star gives 0\farcs18, implying an equivalent gaussian width of $\sim
0\farcs17$ for the arc itself.  Figure 2 shows the brightness profile
along the length of the arc (and integrated across the profile). The
arc centroid position was traced and the flux extracted from the image
as if it were a two-dimensional spectrum.  The error bars are based on
the number of electrons in the image and the observed sky noise. This
figure confirms the visual impression that there is real variation in
surface brightness along the length of the arc and the peak of the
profile occurs at the center, consistent with a normal galaxy
luminosity profile.

In addition to the arc itself, other faint images are evident,
especially in the subtracted image (Fig. 1b).  Large magnification
events such as arcs are generally due to the presence of a source near
the caustic lines (e.g., \citet{Ref94}). In simple lens geometry
configurations (spherical or elliptical) they are usually accompanied
by additional images on the opposite side of the lens.  Models of the
CFRS03.1077 lens indicate that faint images near the position of the
lensing galaxy (visible near the center of the lens in Fig. 1b)  are
likely to be counter-images to the arc.  However, they could also be
small objects embedded within the lensing galaxy, or simply a
projection of foreground/background objects (with the only reservation
that if they are unlensed objects, they should be then at a redshift
significantly lower than that of the arc source). Unfortunately, we do
not have color information of these images. Since their location is
very suggestive that they are counter images of the arc and in the
absence of other data, we assume in the following that they are, in
fact, lensed images.
   
\subsection{CFHT spectroscopy}

A spectrum of the lensing galaxy, CFRS03.1077, from the original CFRS
survey is shown in Figure 3.  Based mostly on the location of the
4000\AA\ break and its overall spectral energy distribution,
\citet{Ham95} assigned a redshift of z = 0.938 with a high confidence
level (class 3, see \citet{LeF95}).  The strong 4000\AA\ break and
absence of any [OII] emission indicates that the galaxy is likely to be
an elliptical. The slit was positioned EW and its width was 1\farcs75,
so part of the arc was in the MOS spectrograph slit and, indeed, the
spectrum of a neighboring object, i.e., the arc, was noted by the CFRS
team but no redshift was derived. After recognizing that this was
likely a lensed galaxy, this spectrum was re-reduced and analysed but
still no redshift determination was possible, partly because the
spectrum of the arc was overwhelmed by that of the bright galaxy.
(This is unlike the case of CFRS14.1311, where \citet{Cra96} were able
to subsequently detect strong emission lines of the background
quasar).

Spectra of the arc alone were obtained with the OSIS spectrograph on
CFHT on 1997 August 26--28. Unfortunately, the seeing was only average
during these exposures (0\farcs5 -- 0\farcs8) and the OSIS fast guiding
system, which would
have improved the image quality somewhat, was not available. A
0\farcs75 slit was aligned along the brightest part of the arc at PA =
20$\arcdeg$ on two of the nights, but a 1\farcs0 slit had to be used during
poorer seeing on 1997 Aug 27.  Two spectra were obtained on 1997 Aug 26,
three on Aug 27 and five on Aug 28. All exposures were 1800s, so the
total exposure is equivalent to five hours. Small shifts of the arc
along the slit were made between exposures to improve the flat fielding
and hence signal-to-noise of the final spectra.  The spectra were
recorded with the STIS 2 CCD with 21 micron pixels which
correspond to 3.4\AA\ per pixel with the V150 grism and 0\farcs14 per
pixel in the spatial dimension. The wavelength resolution, set by the
0\farcs75 slit, is 17\AA\ and the usable wavelength range is 4000 --
9000\AA.

The two dimensional spectra were bias subtracted and flat fielded using
a normalized flat derived from spectra of internal calibration lamps.
Two dimensional sections were then extracted from each image, centered
on the spectrum of the arc. These were shifted to compensate for any
wavelength shifts (determined from the night sky lines), combined and
then a one dimensional spectrum was extracted. The
background-subtracted two-dimensional spectrum was inspected to see if
there was any indication of spatial variations of the spectral features
along the arc, but none was detectable.  The spectrum width is $\sim$14
pixels or $\sim$2\farcs0, consistent with length of the arc (see Figures
1 and 2).

A portion of the spectrum of the arc is shown in Figure 4. A spectrum
of Feige 24 was used to provide flux calibration, although the absolute
calibration is uncertain due to the narrow slit width. The spectrum was
cross-correlated against $HST$ ultraviolet spectra of NGC1741 and
NGC4214, two prototypical starburst galaxies \citep{Con96,
Lei96}. Since no emission lines are obvious in our
spectrum of the arc, the strong Ly $\alpha$ emission lines were first
excised from these starburst spectra. Both of the resulting templates
give a strong cross correlation peak at z = 2.94, the same redshift as
derived from a visual examination of the spectrum of the arc. The rest
wavelengths of the most prominent features familiar in nearby starburst
galaxies are identified in Fig. 4, with a spectrum of NGC4214  shown
for comparison.  In common with many of the z $\sim$ 3 galaxies
observed, for example, by \citet{Ste96} and \citet{Pet97}, 
no Ly $\alpha$ emission is visible, the continuum is flat and
drops shortward of Ly $\alpha$, and numerous stellar and interstellar
lines of C, O, Si and Al are visible. As with all such galaxies
\citep{Pet97}, there is considerable scatter among the
velocities derived from the individual features, and we adopt z = 2.941
with an estimated error of 0.008. We note that, in retrospect, some of
the features identified in Fig. 4 are visible on our original MOS spectrum of
the arc, lending additional weight to our redshift determination.

\section{Lens Modelling}

The lens CFRS03.1077 was modelled assuming an isothermal mass profile.
Gravitational bending angles have been computed using the code AIGLE
(see \citet{Ham97}) for an elliptical lens with an isothermal profile
($\rho = \rho_{0}(1+(r/r_c)^2)^{-1}$) where r$_c$ is the core radius.
Mass distributions are used rather than potentials, to avoid any
further assumptions in mass estimations.  The arc itself can be
perfectly reproduced by the model, assuming a uniform circular source
(radius r$_s$ = 0\farcs1) at z = 2.941, and a lens with parameters
provided by the light distribution (see Table 1). However, since the
true lensing parameter is the mass enclosed within a circle defined by
the arc, there is a clear degeneracy between the core radius and the
line of sight velocity dispersion ($\sigma_{los}$). It is beyond the
scope of this paper to accurately evaluate the available range of
acceptable parameters, but an initial evaluation gives $\sigma_{los}=
390\pm$70km s$^{-1}$, and small core radii (r$_c$ from 0\farcs1 to
0\farcs3 ).  It is more interesting to note that for a large fraction
of the (r$_c$, $\sigma_{los}$) configurations, the model predicts
additional images on the other side of the lensing galaxy. These fall
within 0\farcs03 of the locations of the two faint blobs on the
opposite side of the lensing galaxy (discussed in section 2.1.2) if a
slight offset of the lens position angle, from 104 to 122\arcdeg\ is
assumed. Figure 5a shows a simulation of the three images (I0: the arc,
I1 and I2: the blobs on the other side of the lens, near its center).
Figure 5b shows the location of the source relative to the area of high
magnification.  The source falls at the edge of the elliptical caustic
lines, most of its surface within the three image area. The
magnification factor (10 for the arc alone) provides a magnitude for
the unlensed source of 24.4.

The requirement to properly reproduce this three image configuration
tightly constrains the lensing parameters.  Indeed the actual number
of constraints is 8 (6 for the image positions and two for the
intensity ratios), while the number of parameters is only 6 (r$_c$,
$\sigma_{los}$, ellipticity, P.A. and the source location).  This leads
us to believe that I1 and I2 are likely counter images of the arc, with
the result that the mass estimates of the lens are very accurate.
Actually the lensing model predicts an accuracy as small as 2 km
s$^{-1}$ for the lens velocity dispersion. The cosmological parameter
($q_{0}$ or $\Lambda$) is another factor of uncertainty.  However, the
gravitational bending angle depends on the ratio of the angular
diameter distances between the lens and the source and of the lens and
so the dependency on the cosmological parameters is small. For the
present example, $q_{0}$= 0.5 leads to $\sigma_{los}$= 390$\pm$2 km
s$^{-1}$ and $q_{0}$= 0 to $\sigma_{los}$= 384$\pm$2 km s$^{-1}$. Our
determination of the mass parameter can be summarized by
$\sigma_{los}$= 387$\pm$5 km s$^{-1}$.

\subsection{Evolution of early type galaxies}

The photometric model fitting to the luminosity profile of the lensing
galaxy that was described in section 2.1.1 gives the central surface
brightness $\mu_0$ and half-light radius $R_e$ corresponding to an
$R^{1/4}$ law. These values allow us to derive an expected velocity
dispersion from the fundamental plane relations between size, surface
brightness, and velocity dispersion.  \citet{Jor96} give the required
relations. The observed central surface brightness is $\mu_0=16.20$ in
$I_{AB}$. At a redshift of $z=0.938$ this corresponds to
$\mu_{0,B}(restframe)=\mu_{0,I}-7.5\log(1+z) -0.13$ where the latter
term is the correction from observed wavelength to $B$-band. This
value, $\mu_{0,B}(restframe)=13.92$, converts to $\mu_{0,rgunn}=13.24$
and surface brightness at $R=R_e$, $<\mu_{e}>_r=20.18$. Using the
relations given by \citet{Jor96} yields a velocity dispersion of
$\sigma=380^{+65}_{-60}$. This is in very good agreement with the mass
parameter derived from the lensing model above.  The fact that it
agrees so well, however, poses another problem in that it apparently
represents a contradiction to the passive evolution model for
elliptical galaxies. For example, \citet{Sch99} claim evolution of
nearly 1 magnitude in luminosity (or, equivalently, surface brightness)
for an elliptical at this redshift. In fact, the lensing galaxy is one
of the sample of elliptical galaxies in the CFRS/LDSS imaging survey
that was used to derive the evolution. In order to estimate mass from
the photometric parameters of this galaxy the observed surface
brightness should be ``de-evolved", that is, the central surface
brightness should be made one magnitude fainter before it is compared
to the local ($z=0$) fundamental plane.  If this is done, a mass
estimate of $\sigma=214^{+35}_{-30}$ is obtained, obviously very
significantly lower than the mass estimate from the lens model.

This result for CFRS03.1077 is contrary to results from studies of the
evolution of the fundamental plane at moderate redshift
\citep{kel00,vd01} and of lensing galaxies on average.  \citet{Koc00}
present data from 30 lenses and conclude that the fundamental plane for
both field galaxies and cluster galaxies are similar and that they are
representative of passively evolving populations formed at z $>$ 2.
That view is consistent with the fundamental plane studies. The reason
that CFRS03.1077 does not appear  to share this behavior is not
obvious. Is there substantial variance in the evolution history of
massive elliptical galaxies?  An examination of Figure 3 from
\citet{Sch99} shows that this particular galaxy lies within its
$1\sigma$ error bar of the no evolution locus for elliptical galaxies
in the $M_B(AB)-\log R_e$ plane and is one of the most deviant points
from the evolving relation. The only other very luminous elliptical
galaxy in the study by \citet{Sch99} (CFRS 14.1311 at $z=0.807$) also
appears in Figure 3 of that paper and lies exactly on the evolving
$M_B(AB)-\log R_e$ relation implying an evolution in luminosity (or,
equivalently, in surface brightness) of $\sim 1$ magnitude. Remarkably,
that galaxy is an Einstein Cross lens discovered by \citet{Rat95} and
studied by \citet{Cra96} who report a velocity dispersion of 230 km
s$^{-1}$ as implied by the lens model. If the observed values of $R_e$
and $<\mu>_e$ for CFRS 14.1311 are converted into a velocity dispersion
via the local fundamental plane then a value of $\sigma=537$ km
s$^{-1}$ is obtained. If an evolution of one magnitude in surface
brightness is then applied this is reduced to $\sigma=290$km s$^{-1}$, roughly
consistent with the value derived from the lens model. In other words,
CFRS14.1311 behaves as expected but CFRS03.1077 does not. It should be
noted that neither of these galaxies have detectable levels of
[OII]3727 emission which would indicate star formation and, furthermore,
the galaxy with lower surface brightness (CFRS03.1077) is the bluer of
the two galaxies, opposite to what might be expected if a recent burst of 
star formation was a factor.

In summary, we have two independent methods of estimating the velocity
dispersion of these two lens galaxies. The photometric method uses size
and surface brightness to estimate the expected velocity dispersion
assuming that these galaxies lie on either the local fundamental plane
or on an evolved fundamental plane. The lens method is independent of
the galaxy light and depends only on the lens geometry. The results of
these two methods can be made to be consistent only if CFRS03.1077 lies
on the un-evolved fundamental plane whereas CFRS 14.1311 has evolved by
$\sim 1$ magnitude at $z=0.807$. In other words, the results suggest
that these two highly luminous field elliptical galaxies may have had
very different evolutionary histories.

\section{CONCLUSION}

Our spectroscopic observations confirm that the arc surrounding the z =
0.938 elliptical galaxy CFRS03.1077 is indeed a lensed image of a
background galaxy. The redshift of this galaxy is z = 2.941. Standard
lens models easily reproduce the observed arc structure and also
suggest that two faint objects observed near the lensing galaxy on the
opposite side to the arc are lensed images. Observations at
other wavelengths should be obtained to determine the colors of these
objects. If they are the same as the arc, then this would be further
evidence that they are lensed images and can be confidently used to
constrain the lens geometry. Multi-wavelength observations of
CFRS03.1077 could also be used to examine whether the internal colors
of the galaxy itself are normal or show strong variations indicative of
recent star formation (cf \citet{Men01}). If the
internal colors are not homogeneous this may help explain why the line
of sight velocity dispersion determined from the lens model is higher
than expected from fundamental plane considerations assuming passive
evolution since $z \sim 1$.

CFRS03.1077 demonstrates the potential offered by detailed study of
Einstein ring lenses. With the advent of 8m class telescopes,
especially those equipped with integral field spectrographs, the
spectroscopic data reported here can now be very significantly improved.
Images at other wavelengths should be obtained to establish or identify
additional lensed images in order to more tightly constrain the lens
model.

\begin{acknowledgments}

\end{acknowledgments}


\clearpage

\begin{deluxetable}{rrrrrrr} 
\tablecolumns{7} 
\tablewidth{0pc} 
\tablecaption{Parameters of the lens CFRS03.1077} 
\tablehead{ 
\colhead{R$_e$} &  \colhead{b/a} &   \colhead{PA}   &  \colhead{$I_{AB}$} & 
\colhead{$(V - I)_{AB}$} & \colhead{$M_{AB}(B)$} & \colhead{$(U - V)_{0,AB}$} \\
\colhead{arcsec} &   & \colhead{deg}    & \colhead{mag} & 
\colhead{mag}    & \colhead{mag}   & \colhead{mag}
}
\startdata 

1.74$\pm$0.1 & 0.67$\pm$0.03 & 104.4$\pm$3 & 20.36$\pm$0.05 & 2.13$\pm$0.15
& $-$22.77$\pm$0.1 & 1.53$\pm$0.2 \\ 

\enddata 

\tablenotetext{1}{Assumes H$_o$ = 50 km s$^{-1}$ Mpc$^{-1}$ and q$_0$ = 0.5.}
\end{deluxetable}

\begin{deluxetable}{cccccc} 
\tablecolumns{6} 
\tablewidth{0pc} 
\tablecaption{Parameters of the arc near CFRS03.1077} 
\tablehead{ 
\colhead{z} &  \colhead{Length} &   \colhead{Width}   &  \colhead{SB($I_{AB},max)$} & 
\colhead{I$_{AB}$} & \colhead{M(B)$_{AB}$} \\
\colhead{} &   \colhead{arcsec}    & \colhead{arcsec} & \colhead{mag} &
\colhead{mag}    & \colhead{mag}   
}
\startdata 

2.941 & 2.1 & 0.17$\pm$0.02 & 21.2 & 21.94 & $-$22.5  \\ 

\enddata 

\tablenotetext{1}{Assumes H$_o$ = 50 km s$^{-1}$ Mpc$^{-1}$ and q$_0$ =
0.5. M(B)$_{AB}$ is calculated assuming a lensing magnification factor
of 10. It also assumes a very blue spectral energy distribution and
corresponds to a lower limit on the B-band luminosity. The absolute
magnitude at AB = 2030 angstroms is 0.41 mag fainter.}
\end{deluxetable} 

\clearpage
\begin{figure}
\plotone{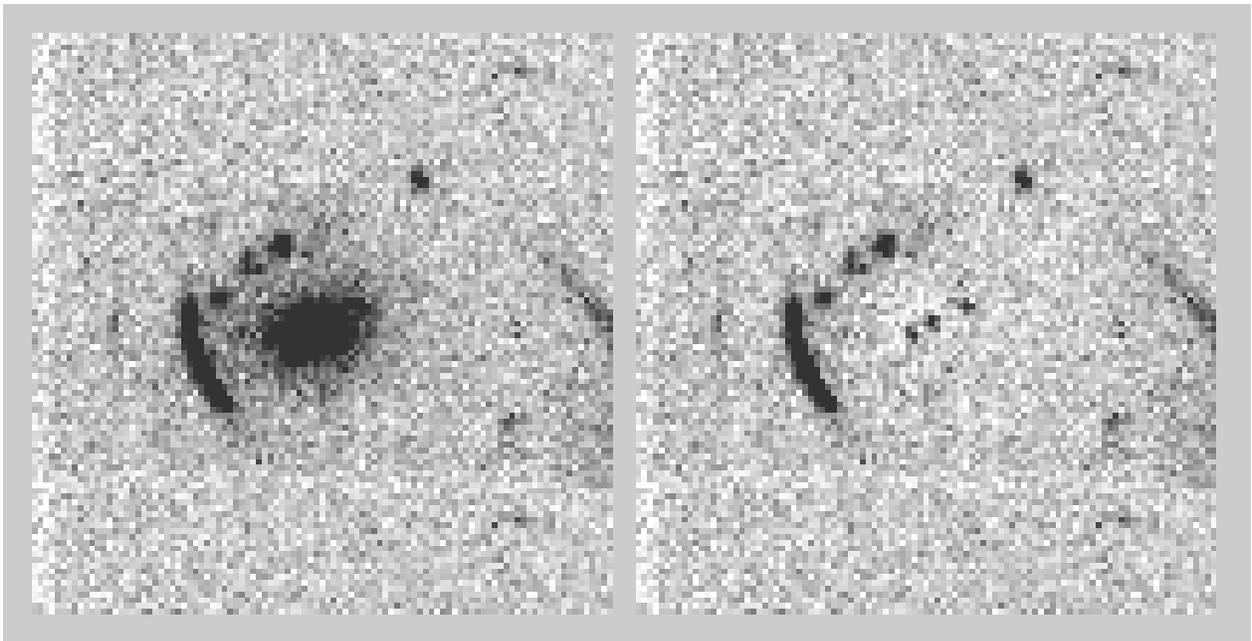}
\caption{(a: left) $HST$ F814W image of CFRS03.1077 showing the
bright arc to the east. (b: right) Same image after the
subtraction of the ``symmetrized" galaxy as discussed in the text. The images are 10\arcsec\ square, N is up and E to the left.
}
\end{figure}

\clearpage

\begin{figure}
\plotone{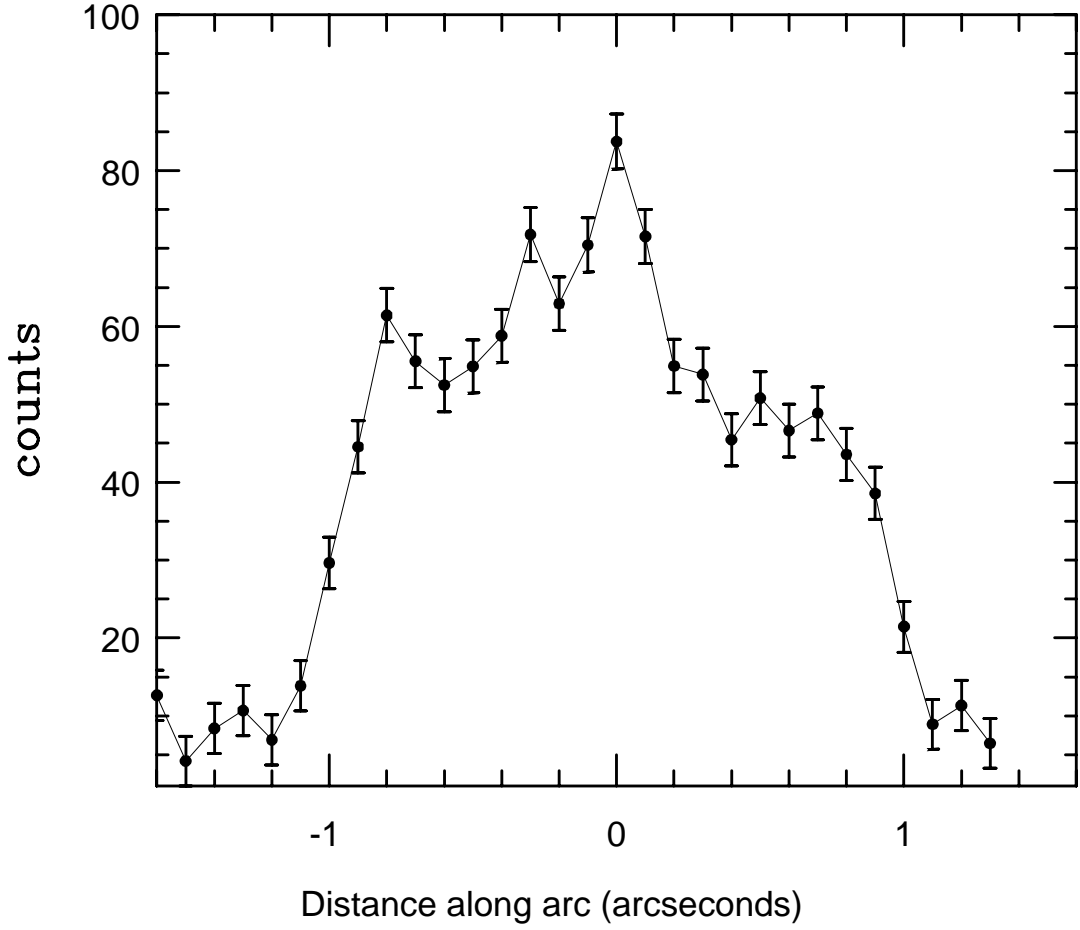}
\caption{Distribution of intensity along the arc (integrated across its width)}
\end{figure}

\begin{figure}
\plotone{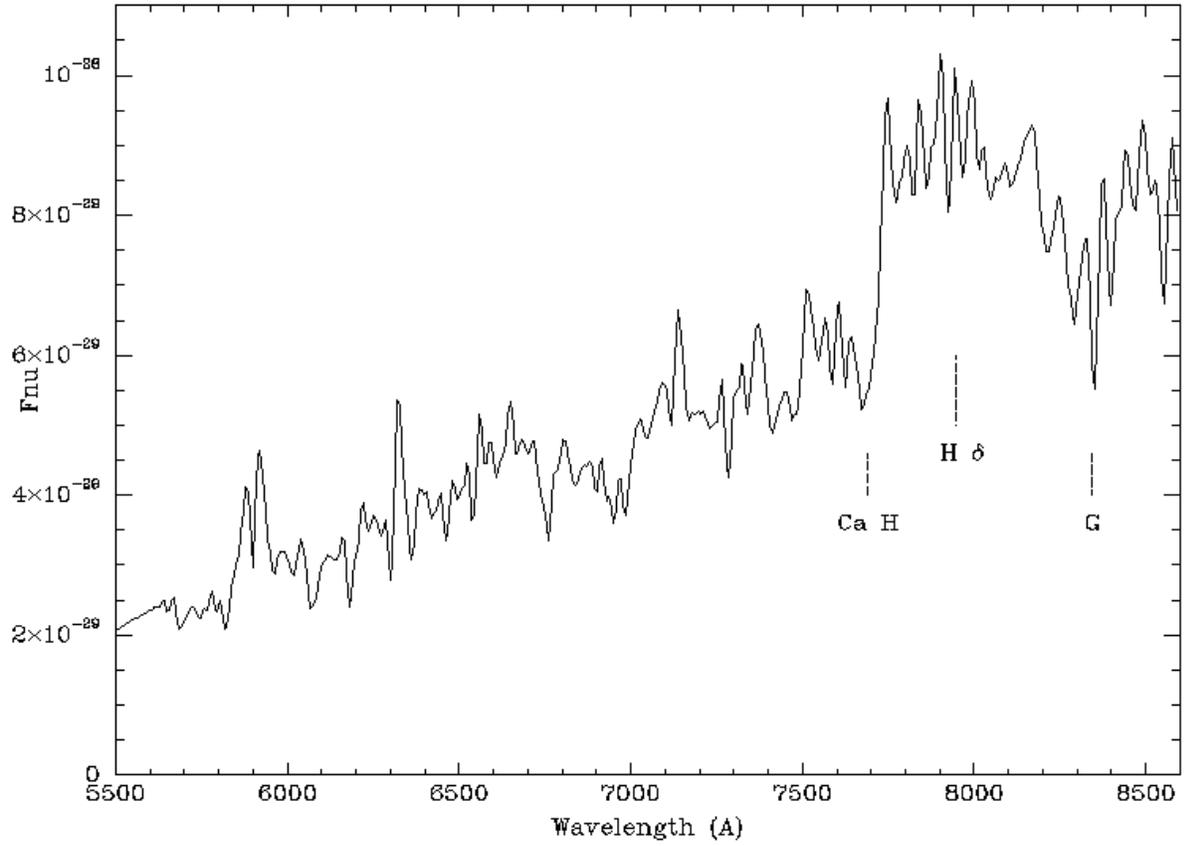}
\caption{CFHT MOS spectrum of the lensing galaxy, CFRS03.1077. The locations of
some typical features redshifted to z = 0.938 are marked, although the large
4000\AA\ break is the most distinctive feature.}
\end{figure}

\begin{figure}
\plotone{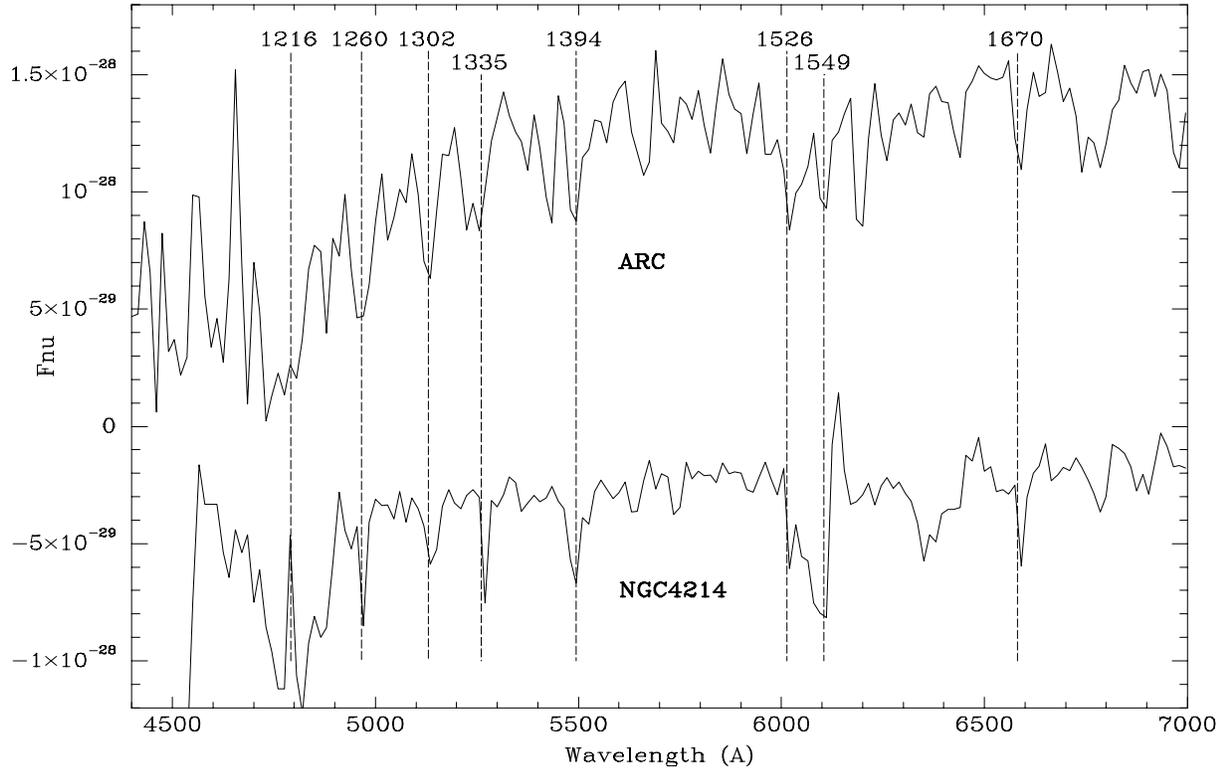}
\caption{The spectrum of the arc (upper)  with a redshifted spectrum of the starburst galaxy NGC4214 (Leitherer et al. 1996) shown for comparison. The flux scale of the latter was arbitrarily scaled and shifted for display purposes. The
dashed lines indicate the nominal positions of common absorption lines assuming z = 2.941 (their restframe wavelengths are indicated at the top of the figure). }
\end{figure}

\begin{figure}
\plottwo{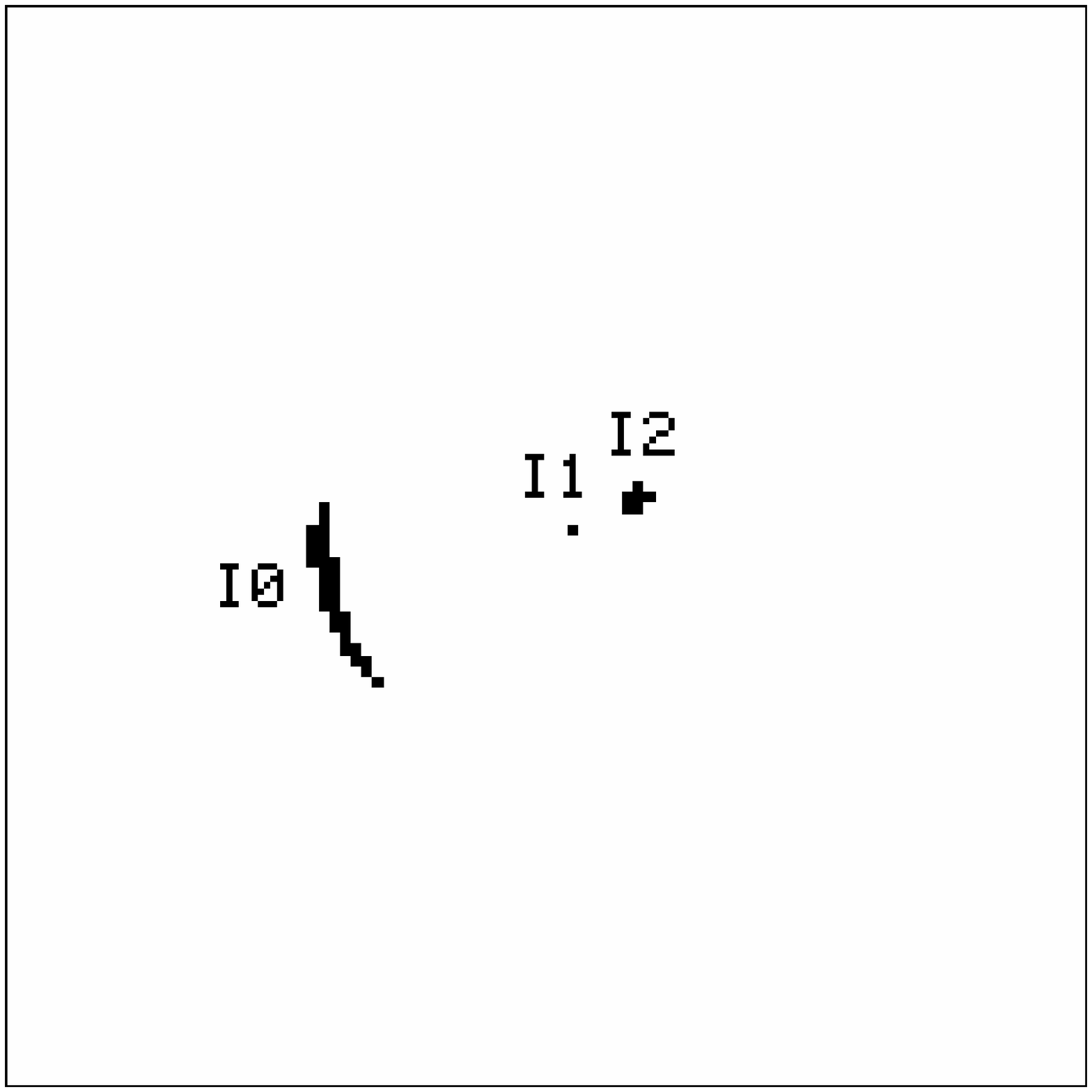}{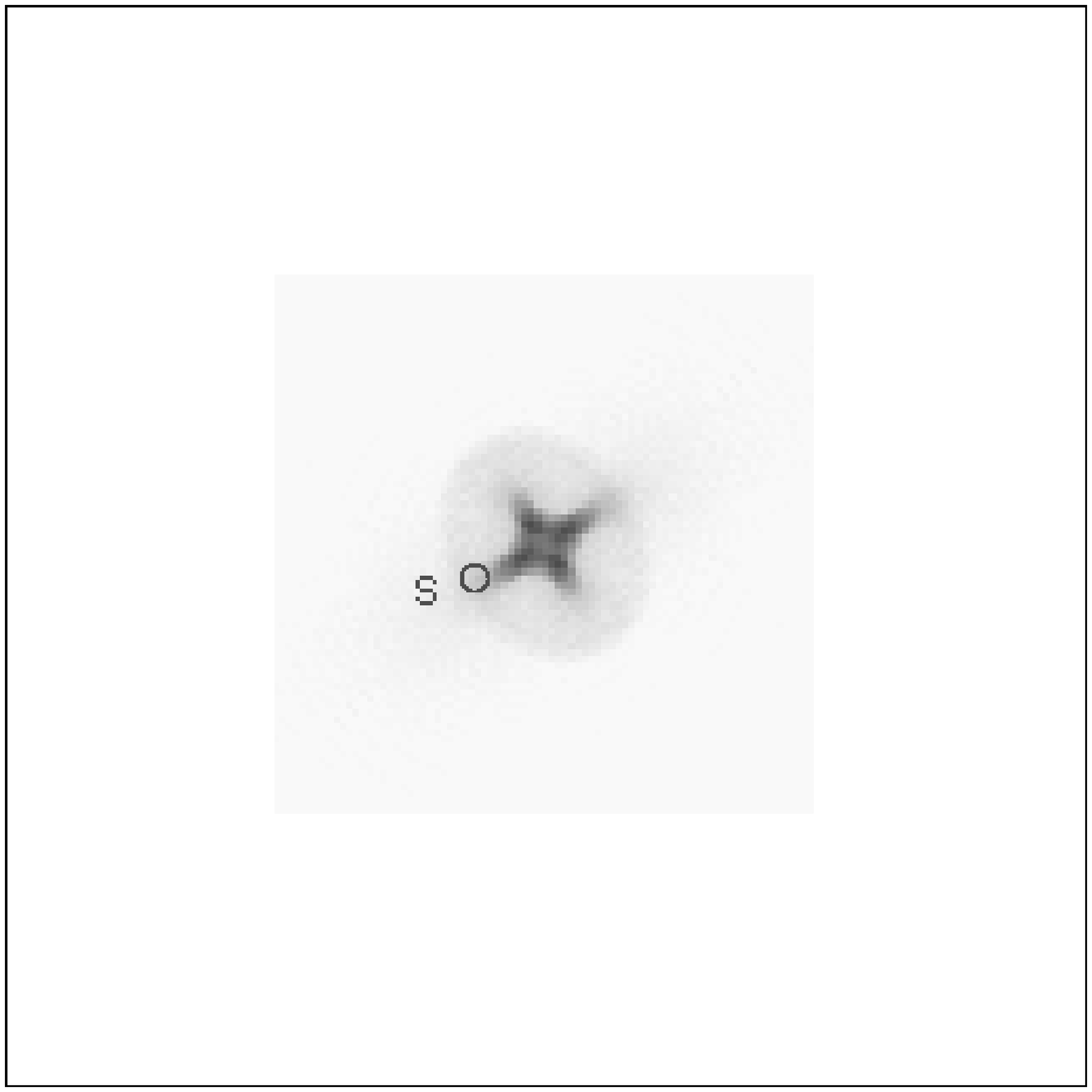}
\caption{
 (a: left) Model of the arc (I0) and the two counter images I1 and
I2. The model assumes $\sigma_{los}=390$ km s$^{-1}$, r$_c$=0\farcs19,
ellipticity = 0.24 and P.A. = 122\arcdeg\ for the lensing galaxy. The
magnification factor is 12, and the ratio between the arc and counter
image luminosities is 0.17. (b: right) High magnification areas
(caustics) in the source plane. This has been computed assuming a
source at z = 2.941 with r = 0\farcs1 (same parameters as for the lens
in Fig. 5a). The faint elliptical boundary delimits the one to three
image areas, and the diamond shape the three to five image areas. A
source with an impact parameter of (x = $-$0\farcs47 and y =
$-$0\farcs2) would provide three images (I0, I1 and I2) in the image
plane as shown in the panel on the left.
}
\end{figure}

\end{document}